\documentclass[prl]{revtex4}
\usepackage{epsfig}

\begin{document}
\title{Two state scattering problem to Multi-channel scattering problem: Analytically solvable model}
\author{Aniruddha Chakraborty \\
School of Basic Sciences, Indian Institute of Technology Mandi,\\
Mandi, Himachal Pradesh 175001, India.}

\begin{abstract}
\noindent Starting from few simple examples we have proposed a general method for finding an exact analytical solution
for the two state scattering problem in presence of a delta function coupling. We have also extended our model to deal with general one dimensional multi-channel scattering problems.
\end{abstract}

\maketitle

\section{Introduction:}
\noindent Nonadiabatic transition due to potential curve crossing is one of the most important mechanisms to
effectively induce electronic transitions in collisions
\cite{Naka1}. This is a very interdisciplinary concept and appears
in various fields of physics and chemistry and even in biology
\cite{Naka2,Naka3,Nikitin,Niki,Child,Book}. The theory of
non-adiabatic transitions dates back to $1932$, when the
pioneering works for curve-crossing and non-crossing were
published by Landau \cite{Landau}, Zener \cite{Zener} and
Stueckelberg \cite{Stueckelberg} and by Rosen and Zener
\cite{Rosen} respectively. Osherov and Voronin solved the case where
two diabatic potentials are constant with exponential coupling
\cite{Voronin}. C. Zhu solved the case where two diabatic
potentials are exponential with exponential coupling
\cite{Nikitinmodel}. In this paper we consider the case of two or more diabatic potentials with Dirac Delta couplings. The Dirac Delta coupling  model has the advantage that it can be exactly solved \cite{AniThesis,AniBook,Ani1,Ani2,Ani3,Ani4} if the uncoupled diabatic potential has an exact solution.

\section{Our model:}

\noindent We consider two diabatic curves, crossing each other. There is a coupling
between the two curves, which cause transitions from one curve to another.
This transition would occur in the vicinity of the crossing point. In
particular, it will occur in a narrow range of x, given by
\begin{equation}
\label{1}V_1(x)-V_2(x)\simeq V_{12}(x_c),
\end{equation}
\noindent where $x$ denotes the nuclear coordinate and $x_c$ is the crossing point. $V_1$ and $V_2$ are determined by the shape of the diabatic curves and $V_{12} $ represents the coupling between them. Therefore it is interesting to
analyse a model where coupling is localized in space near $x_c$. Thus we
put
\begin{equation}
\label{2}V_{12}(x)=K_0\delta (x-x_c),
\end{equation}
\noindent where $K_0$ is a constant. 

\section*{Three simple examples}

We start with a particle moving on any of the two diabatic curves and the
problem is to calculate the probability that the particle will still be on
that diabatic curve after a time $t$. We write the probability amplitude for
the particle as
\begin{equation}
\label{104}\Psi (x)=\left(
\begin{array}{c}
\psi _1(x) \\
\psi _2(x)
\end{array}
\right),
\end{equation}
where $\psi _1(x,t)$ and $\psi _2(x,t)$ are the probability amplitude for
the two states. The Hamiltonian is given by
\begin{equation}
\label{105}H=\left(
\begin{array}{cc}
H_{11}(x) & V_{12}(x) \\
V_{21}(x) & H_{22}(x)
\end{array}
\right),
\end{equation}
where $H_1(x)$, $H_2(x)$ and $V_{12}(x)$ are defined by
\begin{eqnarray}
&H_{11}(x)=-\frac 1{2m}\frac{\partial ^2}{\partial x^2}+V_1(x),\\\nonumber
&H_{22}(x)=-\frac 1{2m}\frac{\partial ^2}{\partial x^2}+V_2(x)\;\; \text{and}\\\nonumber
&V_{12}(x)= V_{21}(x)=K_0\delta (x-x_c).
\end{eqnarray}
The above $V_1(x)$ and $V_2(x)$ are determined by the shape of that
diabatic curve. $V(x)$ is a coupling function which we assume to be a Dirac delta function. The time-independent Schr$\stackrel{..}{o}$dinger equation is writen in the matrix form
\begin{equation}
\label{109}\left(
\begin{array}{cc}
H_{11}(x) & V_{12}(x) \\
V_{21}(x) & H_{22}(x)
\end{array}
\right) \left(
\begin{array}{c}
\psi _1(x) \\
\psi _2(x)
\end{array}
\right) =E\left(
\begin{array}{c}
\psi _1(x) \\
\psi _2(x)
\end{array}
\right).
\end{equation}
This is equivalent to
\begin{eqnarray}
H_{11}(x)\psi _1(x)+K_0\delta (x-x_c)\psi _2(x)=E\psi _1(x)\;\;\; \text{and} \\\nonumber
K_0\delta (x-x_c)\psi _1(x)+H_{22}(x)\psi _2(x)=E\psi _2(x).
\end{eqnarray}
Integrating the above two equations from $x_c-\eta $ to $x_c+\eta $ (where $\eta \rightarrow 0$) we get we get the following two boundary conditions
\begin{eqnarray}
-\frac{\hbar ^2}{2m}\left[\frac{d\psi _1(x)}{dx}\right]_{x_c-\eta
}^{x_c+\eta }+K_0\psi _2(x_c)=0\;\;\; \text{and} \\\nonumber
-\frac{\hbar ^2}{2m}\left[\frac{d\psi _2(x)}{dx}\right]_{x_c-\eta
}^{x_c+\eta }+K_0\psi _1(x_c)=0.
\end{eqnarray}
Also we have two more boundary condition
\begin{eqnarray}
\psi _1(x_c-\eta )=\psi _1(x_c+\eta)\;\;\;\; \text{and}\\\nonumber
\psi _2(x_c-\eta )=\psi _2(x_c+\eta ).
\end{eqnarray}
Using the above four boundary conditions we derive the transition
probability from one diabatic potential to the other, in the case of coupling between (a) two constant potentials, (b) two linear potentials and (c) two exponential potential.

\subsubsection*{Exact analytical solution for constant potential case:}
\noindent In region 1 $(x<x_c)$, the time-independent Schr$\stackrel{..}{o}$dinger
equation for the first potential is given by
\begin{equation}
\label{116}-\frac 1{2m}\frac{\partial ^2\psi _1(x)}{\partial x^2}=E\psi
_1(x).
\end{equation}
The above equation has the following solution
\begin{equation}
\label{117}\psi _1(x)=Ae^{ik_1x}+Be^{-ik_1x},
\end{equation}
where $k_1=\sqrt{\frac{2m}{\hbar ^2}E}$. In region 2 $(x>x_c)$, the time-independent Schr$\stackrel{..}{o}$dinger equation for the first potential is given by
\begin{equation}
\label{118}-\frac 1{2m}\frac{\partial ^2\psi _1(x)}{\partial x^2}=E \psi _1(x)
_1(x)
\end{equation}
Physically acceptable solution of the above equation is given by
\begin{equation}
\label{119}\psi _1(x)=Ce^{ik_1x}.
\end{equation}
In region 1 $(x<x_c)$, the time
independent Schr$\stackrel{..}{o}$dinger equation for the second potential is given by
\begin{equation}
\label{120}-\frac 1{2m}\frac{\partial ^2\psi _2(x)}{\partial x^2}+V_2\psi
_2(x)=E\psi _2(x).
\end{equation}
Physically acceptable solution is given by
\begin{equation}
\label{121}\psi _2(x)=De^{-ik_2x},
\end{equation}
where $k_2=\sqrt{\frac{2m}{\hbar ^2}(E-V_2)}$. In region 2 $(x>x_c)$, the time-independent Schr$\stackrel{..}{o}$dinger
equation for the second potential is given by
\begin{equation}
\label{122}-\frac 1{2m}\frac{\partial ^2\psi _2(x)}{\partial x^2}+V_2\psi
_2(x)=E\psi _2(x).
\end{equation}
Physically acceptable solution is given by
\begin{equation}
\label{123}\psi _2(x)=Fe^{ik_2x}.
\end{equation}
Here we put $x_c=0$. Now using four boundary conditions we calculate
\begin{equation}
\label{124}\left| \frac FA\right| ^2=\left| \frac{imK_0k_1\hbar ^2}{%
(k_1k_2\hbar ^4+m^2K_0^2)}\right| ^2.
\end{equation}
and
\begin{equation}
\label{125}\left| \frac DA\right| ^2=\left| \frac{imK_0k_1\hbar ^2}{%
(k_1k_2\hbar ^4+m^2K_0^2)}\right| ^2.
\end{equation}
So, the transition probability is given by
\begin{equation}
\label{126}T=2\frac{k_2}{k_1}\left| \frac{imK_0k_1\hbar ^2}{(k_1k_2\hbar
^4+m^2K_0^2)}\right| ^2.
\end{equation}
\begin{figure}
\centering \epsfig{file=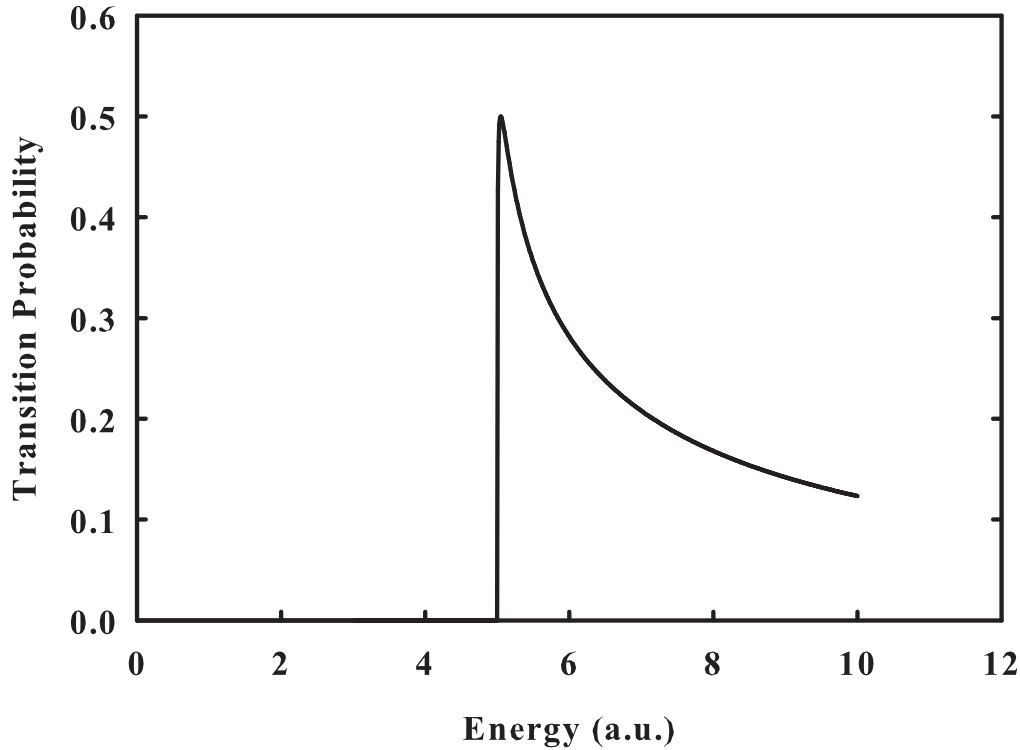,width=0.8\linewidth} \caption{
The plot of transition probability from one constant potential to another constant potential as a function of
energy of incident particle ($K_0$=1.0).} \label{Const}
\end{figure}
\noindent In our numerical calculation we use atomic units so that $\hbar=1$. In the atomic units, we set $V_1(x)=0$, $V_2(x)=5$, $K_0=1.0$ and $m=1.0$. The result of our calculation is shown
in Fig. \ref{Const}.

\subsubsection*{Exact analytical solution for Linear potential case:}

\begin{figure} \centering
\epsfig{file=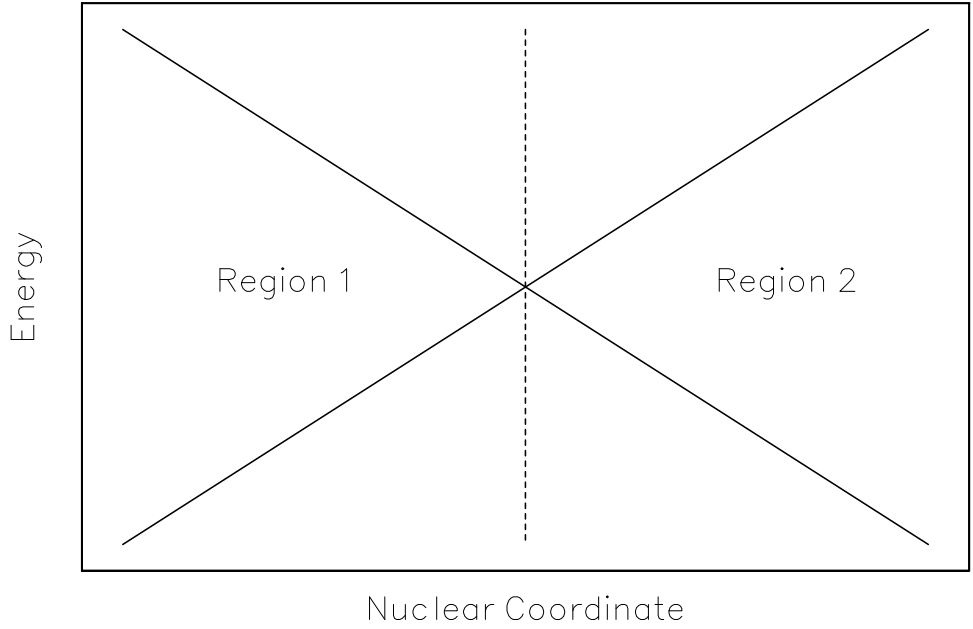,width=0.8\linewidth} 
\caption{Schematic diagram of the two state problem, where one linear potential is coupled to another linear potential in the diabatic representation.} \label{fig1}
\end{figure}

The time-independent Schr$\stackrel{..}{o}$dinger equation for the case
where a linear potential coupled to another linear potential through a Dirac
delta interaction is given below (see Fig. \ref{fig1}).
\begin{equation}
\label{127}\left(
\begin{array}{cc}
-\frac 1{2m}\frac{\partial ^2}{\partial x^2}+p_1x & K_0\delta (x-x_c) \\
K_0\delta (x-x_c) & -\frac 1{2m}\frac{\partial ^2}{\partial x^2}-p_2x
\end{array}
\right) \left(
\begin{array}{c}
\psi _1(x) \\
\psi _2(x)
\end{array}
\right) =E\left(
\begin{array}{c}
\psi _1(x) \\
\psi _2(x)
\end{array}
\right).
\end{equation}
The Eq. (\ref{127}) can be split into two equations
\begin{eqnarray}
\left(-\frac 1{2m}\frac{\partial ^2}{\partial x^2}+p_1x\right)\psi
_1(x)+K_0\delta (x)\psi _2(x)=E\psi _1(x)\;\;\; \text{and}\\\nonumber
\left(-\frac 1{2m}\frac{\partial ^2}{\partial x^2}-p_2x\right)\psi
_2(x)+K_0\delta (x)\psi _1(x)=E\psi _2(x).
\end{eqnarray}
In our calculation, we use $p_{1}=p_{2}=1$. The Time-independent Schr$\stackrel{..}{o}$dinger equation for the first diabatic potential is given below,
\begin{equation}
\left(-\frac 1{2m}\frac{\partial ^2}{\partial x^2}+x\right)\psi _1(x)=E\psi
_1(x).
\end{equation}
In region 1 ( $x<x_c$) the physically acceptable solution is given below
\begin{equation}
\label{131}\psi _1(x)=(A+B)A_i\left[2^{\frac 13}(-E+x)\right]+i(A-B)A_i\left[2^{\frac
13}(-E+x)\right].
\end{equation}
Here $A_i[z]$ and $B_i[z]$  represent the Airy functions. In the above expression $A$ denotes the probability amplitude for motion along the positive direction and $B$ denotes the probability amplitude for motion along the negative direction \cite{Ani5}. The physically acceptable solution in region 2 ( $x>x_c$), is given by
\begin{equation}
\label{132}\psi _1(x)=CA_i\left[2^{\frac 13}(-E+x)\right].
\end{equation}
In this region the net flux is zero \cite{Ani5}. \par
\begin{figure}
\centering \epsfig{file=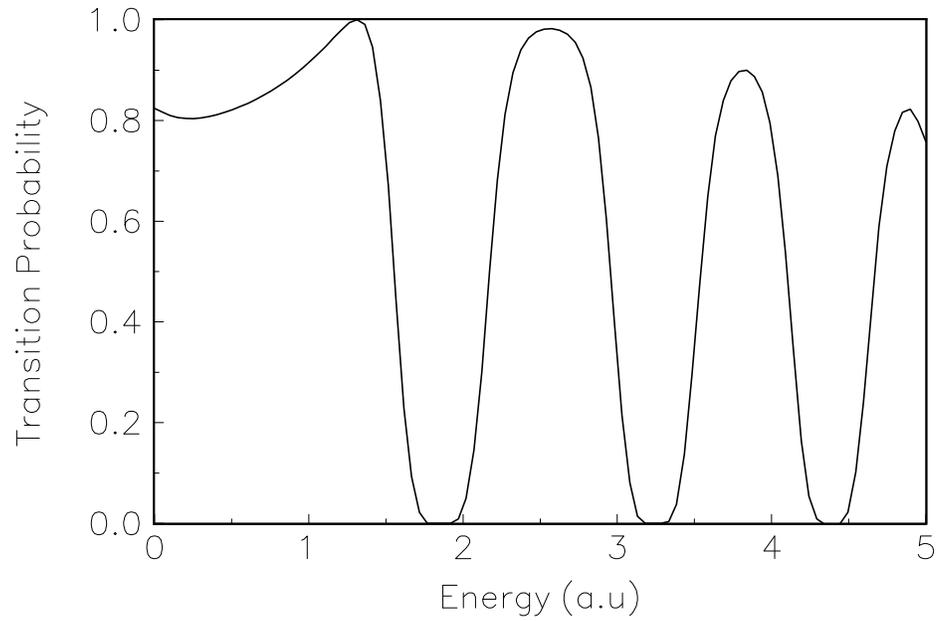,width=0.8\linewidth} \caption{
The plot of transition probability from one linear potential to another linear potential, as a function of
energy of incident particle ($K_0$=1.0).} \label{linear}
\end{figure}

 The time-independent Schr$\stackrel{..}{o}$
dinger equation for the second diabatic potential is given below
\begin{equation}
\label{133}\left(-\frac 1{2m}\frac{\partial ^2}{\partial x^2}-x\right)\psi _2(x)=E\psi
_2(x).
\end{equation}
In region1 ( $x<x_c$) the physically acceptable solution is
\begin{equation}
\label{134}\psi _2(x)=DA_i\left[2^{\frac 13}(-E-x)\right].
\end{equation}
In this region the net flux is zero \cite{Ani5}. In region 2 ( $x>x_c$) the physically acceptable
solution is
\begin{equation}
\label{135}\psi _2(x)=F\left(A_i\left[2^{\frac 13}(-E-x)\right]-iB_i\left[2^{\frac 13}(-E-x)\right]\right).
\end{equation}
Using the four boundary conditions mentioned above (here we put $x_c=0$), we
have derived an analytical expression  for the transition probability from one diabatic potential to the other diabatic potential and the final expression is given below
\begin{equation}
T= 16 \times 2^{2/3}\left| \frac{N}{D} \right|^2,
\end{equation}
where
\begin{eqnarray}
N = Ai\left[-2^{1/3} E \right]^2 \left(- Ai'\left[-2^{1/3} E \right] Bi \left[-2^{1/3} E \right]+ Ai \left[-2^{1/3} E \right] Bi'\left[-2^{1/3} E \right]\right)
\end{eqnarray}
and
\begin{eqnarray}
D = - 8 i Ai\left[-2^{1/3} E\right]^3 Bi\left[-2^{1/3} E\right] +2^{2/3} Ai'\left[-2^{1/3} E\right]^2 Bi\left[-2^{1/3} E\right]^2 \\ \nonumber + 4 Ai\left[-2^{1/3} E\right]^4 -2 2^{2/3} Ai\left[-2^{1/3} E\right] Ai'\left[-2^{1/3} E\right] 
Bi\left[-2^{1/3} E\right] Bi' \left[-2^{1/3} E\right]\\ \nonumber +Ai\left[-2^{1/3} E\right]^2 \left(-4 Bi\left[-2^{1/3} E\right]^2+2^{2/3} Bi'\left[-2^{1/3}
E\right]^2\right).
\end{eqnarray}
In our numerical calculation we set $p_{1}=1$, $p_{2}=1$, $K_0=1$ and $m=1$ in atomic units. The result of our calculation is shown
in Fig. \ref{linear}.

\subsubsection*{Exact analytical solution for exponential potential case:}
We start with the time-independent Schr$\stackrel{..}{o}$dinger equation for
a two state system
\begin{equation}
\label{137}\left(
\begin{array}{cc}
-\frac 1{2m}\frac{\partial ^2}{\partial x^2}+V_0e^{ax} & K_0\delta (x-x_c)
\\
K_0\delta (x-x_c) & -\frac 1{2m}\frac{\partial ^2}{\partial x^2}+V_0e^{-ax}
\end{array}
\right) \left(
\begin{array}{c}
\psi _1(x) \\
\psi _2(x)
\end{array}
\right) =E\left(
\begin{array}{c}
\psi _1(x) \\
\psi _2(x)
\end{array}
\right).
\end{equation}
Eq. (\ref{137}) can be split into the following two equations
\begin{eqnarray}
\label{16}\left(-\frac 1{2m}\frac{\partial ^2}{\partial x^2}+V_0e^{ax}\right)\psi
_1(x)+K_0\delta (x-x_c)\psi _2(x)=E\psi _1(x)\;\;\; \text{and} \\\nonumber
\label{138}\left(-\frac 1{2m}\frac{\partial ^2}{\partial x^2}+V_0e^{-ax}\right)\psi
_2(x)+K_0\delta (x-x_c)\psi _1(x)=E\psi _2(x).
\end{eqnarray}
In our calculation we took $V_0=1.0$ and $a=1$. Using the time-independent Schr$\stackrel{..}{o}$dinger equation the first diabatic potential is given below
\begin{figure} \centering
\epsfig{file=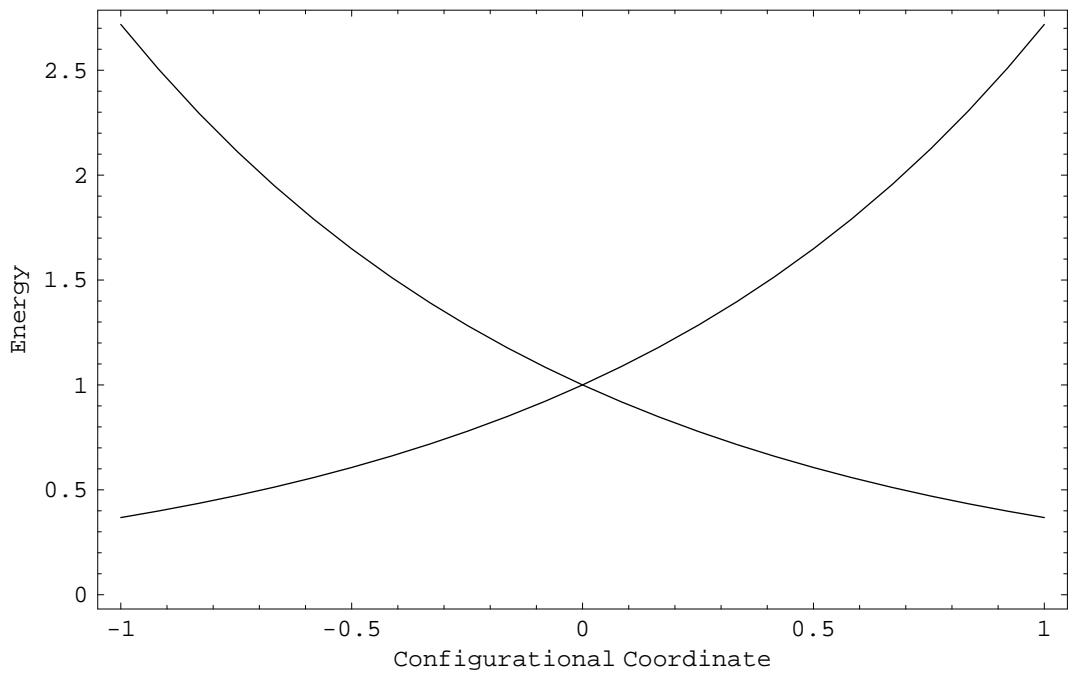,width=0.8\linewidth} 
\caption{Schematic diagram of the two state problem, where one exponential potential is coupled to another exponential potential in the diabatic representation.} \label{exponen}
\end{figure}
\begin{equation}
\label{139}\left(-\frac 1{2m}\frac{\partial ^2}{\partial x^2}+V_0e^{ax}\right)\psi
_1(x)=E\psi _1(x).
\end{equation}
In the region $x<x_c$, the solution of the above equation is given below
\begin{equation}
\label{140}\psi _1(x)=A\,I_{(2\,i\,\sqrt{2E}\,)}(2\,%
\sqrt{2e^x})\,+B\,I_{(-2\,i\,\sqrt{2E}\,)}(2\,\sqrt{2e^x})\,.
\end{equation}
Here $I_{n}(z)$ represent the modified Bessel function of the first kind.
In the above expression $A$ denotes the probability amplitude for motion
along the positive direction and $B$ denotes the probability amplitude for
motion along the negative direction \cite{Ani5}.
In the region, where $x>x_c$, the physically acceptable solution is
\begin{equation}
\label{141}\psi _1(x)=CK_{(-2\,i\,\sqrt{2E}\,)}(2\,\sqrt{2e^x}).
\end{equation}
Here $K_{n}(z)$ represent the modified Bessel function of the second kind. In this region the net flux is zero \cite{Ani3}. \par
\begin{figure}
\centering \epsfig{file=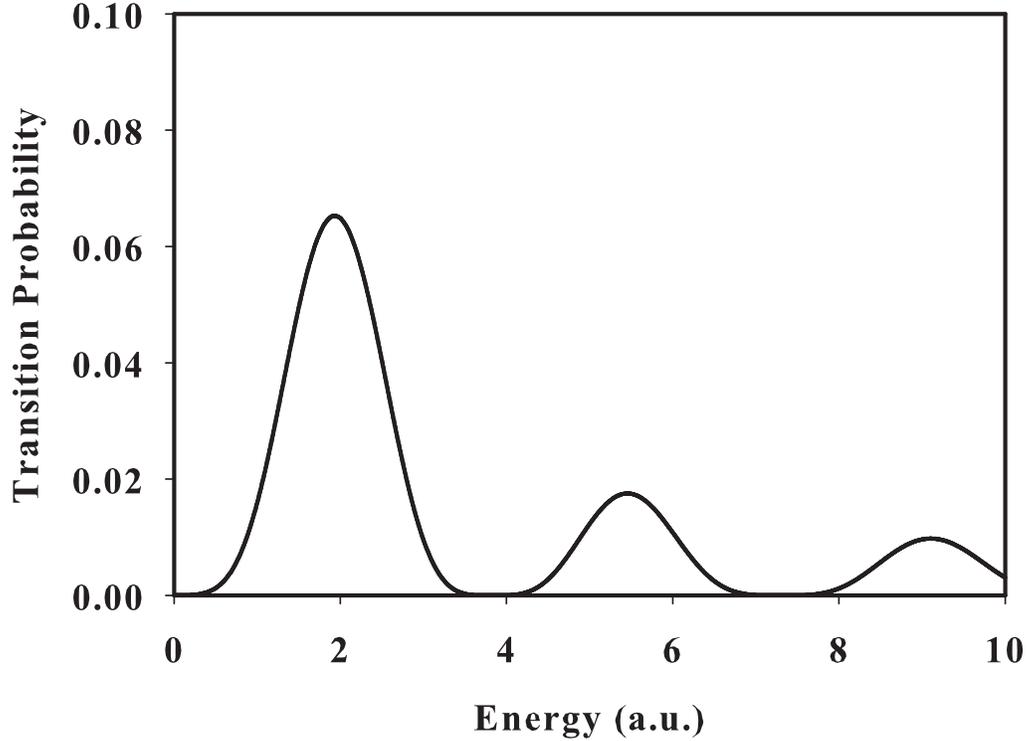,width=0.8\linewidth} \caption{
The plot of transition probability from one exponential potential to the other as a function of
energy of incident particle ($K_0$=0.1).} \label{expo}
\end{figure}

For the second diabatic potential, the  time-independent Schr$\stackrel{..}{o}$dinger equation
is given below
\begin{equation}
\label{142}\left(-\frac 1{2m}\frac{\partial ^2}{\partial x^2}+V_0e^{-ax}\right)\psi
_2(x)=E\psi _2(x).
\end{equation}
In the region, where $x>x_c$, the physically acceptable solution is
\begin{equation}
\label{143}\psi _2(x)=F\,I_{(-2\,i\,\sqrt{2E}\,)}(2\,
\sqrt{2e^{-x}}).
\end{equation}
In the region, where $x<x_c$, the physically acceptable solution is
\begin{equation}
\label{144}\psi _2(x)=DK_{(-2\,i\,\sqrt{2E}\,)}(2\,\sqrt{2e^x}).
\end{equation}
In this region the net flux is zero \cite{Ani3}. 
Using four boundary conditions and putting $x_c=0$, as mentioned before, we
derive an expression for transition probability from one exponential potential to the other exponential potential.
In our numerical calculation we use atomic unit. The value of $K_0=0.1$ and $m=1$. The result of our
calculation is shown in Fig. \ref{expo}.

\section{Formulation of general solution for two state scattering problem using our model}

\noindent We start with the time-independent Schr$\stackrel{..}{o}$dinger equation for
a two state system
\begin{equation}
\label{87}\left(
\begin{array}{cc}
H_{11}(x) & V_{12}(x) \\
V_{21}(x) & H_{22}(x)
\end{array}
\right) \left(
\begin{array}{c}
\psi _1(x) \\
\psi _2(x)
\end{array}
\right) =E\left(
\begin{array}{c}
\psi _1(x) \\
\psi _2(x)
\end{array}
\right).
\end{equation}
\noindent This equation can be written in the following form
\begin{eqnarray}
\psi _2(x)=V_{12}(x)^{-1}\left[E-H_{11}(x)\right]\psi _1(x)\;\;\; \text{and}\\\nonumber
\psi _2(x)=\left[E-H_{22}(x)\right]^{-1}V_{21}(x)\psi _1(x).
\end{eqnarray}
\noindent Eliminating $\psi _2(x)$ from the above two equations we get
\begin{equation}
\label{90}\left[E-H_{11}(x)\right]\psi _1(x) - V_{12}(x)\left[E-H_{22}(x)\right]^{-1}V_{21}(x)\psi
_1(x)=0.
\end{equation}
\noindent The above equations simplify considerably if $V_{12}(x)$ and $V_{21}(x)$ are
Dirac Delta function at $x_c$, which in operator notation may be written as $V=$
$K_0S=K_0$ $|x_c\rangle $ $\langle x_c|$. The above equation now become
\begin{equation}
\label{91}\left(\left[E-H_{11}(x)\right]-K_0^2|x_c\rangle \langle
x_c|\left[E-H_{22}(x)\right]^{-1}|x_c\rangle \langle x_c|\right)\psi _1(x)=0.
\end{equation}
\noindent This may be written as
\begin{equation}
\label{92}\left[H_{11}(x)+K_0^2\delta (x-x_c)G_2^0(x_c,x_c;E)\right]\psi _1(x)=E\psi
_1(x).
\end{equation}
\noindent The above equation can be written in the following form
\begin{equation}
\label{93}\left[E-H_{11}\right]\psi _1(x)=K_0^2\delta (x-x_c)G_2^0(x_c,x_c;E)\psi _1(x),
\end{equation}
\noindent where the right hand side is considered as an inhomogeneous term. The
general solution of this equation can be written as
\begin{equation}
\label{94}\psi _1(x)=\psi _0(x)+K_0^2G_1^0(x,x_c;E)G_2^0(x_c,x_c;E)\psi
_1(x_c),
\end{equation}
\noindent where $\psi _0(x)$ is a solution of the homogeneous equation
\begin{eqnarray}
\left[E-H_{11}\right]\psi _0(x)=0\;\;\; \text{and}\\\nonumber
\left[E-H_{11}\right]G_1^0(x,x^{\prime };E)=\delta (x-x^{\prime}).
\end{eqnarray}
\noindent In the above equation
\begin{eqnarray}
\psi _1(x)=\int\limits_{-\infty }^\infty dx^{\prime }G(x,x^{\prime
};E)\psi _0(x^{\prime })\;\;\; \text{and} \\\nonumber
\psi _0(x)=\int\limits_{-\infty }^\infty dx^{\prime
}G_1^0(x,x^{\prime };E)\psi _0(x^{\prime}).
\end{eqnarray}
\noindent So Eq. (\ref{94}) can be written as
\begin{equation}
\label{99}
\begin{array}{c}
\int\limits_{-\infty }^\infty dx^{\prime }G(x,x^{\prime };E)\psi
_0(x^{\prime })=\int\limits_{-\infty }^\infty dx^{\prime }G_1^0(x,x^{\prime
};E)\psi _0(x^{\prime }) \\
+K_0^2G_1^0(x,x_c;E)G_2^0(x_c,x_c;E)\int\limits_{-\infty }^\infty dx^{\prime
}G(x_c,x^{\prime };E)\psi _0(x^{\prime }).
\end{array}
\end{equation}
\noindent The solution in terms of Green's function is as follows
\begin{equation}
\label{100}G(x,x^{\prime };E)=G_1^0(x,x^{\prime
};E)+K_0^2G_1^0(x,x_c;E)G_2^0(x_c,x_c;E)G(x_c,x^{\prime };E).
\end{equation}
\noindent In the above equation we put $x=x_c$
\begin{equation}
\label{101}G(x_c,x^{\prime };E)=G_1^0(x_c,x^{\prime
};E)+K_0^2G_1^0(x_c,x_c;E)G_2^0(x_c,x_c;E)G(x_c,x^{\prime };E).
\end{equation}
\noindent After simplification, we get
\begin{equation}
\label{102}G(x_c,x^{\prime };E)=\frac{G_1^0(x_c,x^{\prime };E)}{%
1-K_0^2G_1^0(x_c,x_c;E)G_2^0(x_c,x_c;E)},
\end{equation}
\noindent so that
\begin{equation}
\label{103}G(x,x^{\prime };E)=G_1^0(x,x^{\prime };E)+\frac{%
K_0^2G_1^0(x,x_c;E)G_2^0(x_c,x_c;E)G_1^0(x_c,x^{\prime };E)}{%
1-K_0^2G_1^0(x_c,x_c;E)G_2^0(x_c,x_c;E)}.
\end{equation}
\noindent In the above expression, we have the Green's function for two state scattering problem using delta function coupling model. Using the expression of $G(x,x^{\prime };E)$ one can calculate wave function and from the wave function one can easily calculate the transition probability from one diabatic potential to the other.  

\section*{Three channel scattering problem using our model}

We start with a particle moving on any of the three diabatic curves and the
problem is to calculate the probability that the particle will still be in
that diabatic curve after a time $t$. We write the probability amplitude for
the particle as
\begin{equation}
\label{174}\Psi (x)=\left(
\begin{array}{c}
\psi_1(x) \\
\psi_2(x) \\
\psi_3(x)
\end{array}
\right).
\end{equation}
Where $\psi _1(x,t)$, $\psi _2(x,t)$ and $\psi _3(x)$ are the probability
amplitude for the three states. The Hamiltonian matrix of this system is given by
\begin{equation}
\label{175}H=\left(
\begin{array}{ccc}
H_{11}(x) & V_{12}(x) & V_{13}(x) \\
V_{21}(x) & H_{22}(x) & 0 \\
V_{31}(x) & 0 & H_{33}(x)
\end{array}
\right),
\end{equation}
where $H_1(x)$, $H_2(x)$, $H_3(x)$, $V_{12}(x)$, $V_{21}(x)$, $V_{31}(x)$ and $V_{13}(x)$ are
defined by
\begin{eqnarray}
H_{11}(x)=-\frac 1{2m}\frac{\partial ^2}{\partial x^2}+V_1(x),\\\nonumber
H_{22}(x)=-\frac 1{2m}\frac{\partial ^2}{\partial x^2}+V_2(x),\\\nonumber
H_{33}(x)=-\frac 1{2m}\frac{\partial ^2}{\partial x^2}+V_3(x),\\\nonumber
V_{12}(x)=V_{21}(x)=K_0\delta (x-x_c)\;\;\; \text{and}\\\nonumber
V_{13}(x)=V_{31}(x)=K_0\delta (x-x_c).\\\nonumber
\end{eqnarray}
In the above equation $V_1(x)$, $V_2(x)$ and $V_3(x)$ are determined by the shape of that
diabatic curve. The time-independent Schr$\stackrel{..}{o}$dinger equation for this problem is given by
\begin{equation}
\label{181}\left(
\begin{array}{ccc}
H_{11}(x) & K_0\delta (x-x_c) & K_0\delta (x-x_c) \\
K_0\delta (x-x_c) & H_{22}(x) & 0 \\
K_0\delta (x-x_c) & 0 & H_{33}(x)
\end{array}
\right) \left(
\begin{array}{c}
\psi_1(x) \\
\psi_2(x) \\
\psi_3(x)
\end{array}
\right) =E\left(
\begin{array}{c}
\psi_1(x) \\
\psi_2(x) \\
\psi_3(x)
\end{array}
\right).
\end{equation}
This matrix representation is equivalent to the following three equations
\begin{eqnarray}
H_{11}(x)\psi _1(x)+K_0\delta (x-x_c)\psi _2(x)+K_0\delta
(x-x_c)\psi _3(x)=E\psi _1(x),\\\nonumber
K_0\delta (x-x_c)\psi _1(x)+H_{22}(x)\psi _2(x)=E\psi _2(x) \;\; \text{and}\\\nonumber
K_0\delta (x-x_c)\psi _1(x)+H_{33}(x)\psi _3(x)=E\psi _3(x).
\end{eqnarray}
Integrating the above three equations from $x_c-\eta $ to $x_c+\eta$ (where $\eta \rightarrow 0$) we get the following three boundary conditions
\begin{eqnarray}
\label{185}-\frac{\hbar ^2}{2m}\left[\frac{d\psi _1(x)}{dx}\right]_{x_c-\eta
}^{x_c+\eta }+K_0\psi _2(x_c)+K_0\psi _3(x_c)=0,\\\nonumber
-\frac{\hbar ^2}{2m}\left[\frac{d\psi _2(x)}{dx}\right]_{x_c-\eta
}^{x_c+\eta }+K_0\psi _1(x_c)=0\;\; \text{and}\\\nonumber
-\frac{\hbar ^2}{2m}\left[\frac{d\psi _3(x)}{dx}\right]_{x_c-\eta
}^{x_c+\eta }+K_0\psi _1(x_c)=0.
\end{eqnarray}
Also we have three more boundary conditions
\begin{eqnarray}
\psi_1(x_c-\eta )=\psi_1(x_c+\eta )\\\nonumber
\psi_2(x_c-\eta )=\psi_2(x_c+\eta )\;\; \text{and}\\\nonumber
\psi_3(x_c-\eta )=\psi_3(x_c+\eta ).
\end{eqnarray}
\noindent Using the above six boundary conditions we derive analytical expressions for transition
probability from one diabatic potential to the other in the case of coupling between (a) three constant potentials, (b) three linear potentials and (c) three exponential potentials.

\section{Formulation of general solution for multi-channel scattering problem using our model}

\noindent We start with the time-independent Schr$\stackrel{..}{o}$dinger equation for
a three state system, given by
\begin{equation}
\label{146}\left(
\begin{array}{ccc}
H_{11}(x) & V_{12}(x) & V_{13}(x) \\
V_{21}(x) & H_{22}(x) & 0 \\
V_{31}(x) & 0 & H_{33}(x)
\end{array}
\right) \left(
\begin{array}{c}
\psi _1(x) \\
\psi _2(x) \\
\psi _3(x)
\end{array}
\right) =E\left(
\begin{array}{c}
\psi _1(x) \\
\psi _2(x) \\
\psi _3(x)
\end{array}
\right),
\end{equation}
\noindent where
\begin{eqnarray}
H_{11}(x)=-\frac 1{2m}\frac{\partial ^2}{\partial x^2}+V_1(x),\\\nonumber
H_{22}(x)=-\frac 1{2m}\frac{\partial ^2}{\partial x^2}+V_2(x)\;\; \text{and}\\\nonumber
H_{33}(x)=-\frac 1{2m}\frac{\partial ^2}{\partial x^2}+V_3(x).
\end{eqnarray}
\noindent This above matrix equation can be written in the following form
\begin{eqnarray}
\left[H_{11}(x)-E\right]\psi_1(x)+V_{12}(x)\psi_2(x)+V_{13}(x)\psi _3(x)=0,\\\nonumber
\left[H_{22}(x)-E\right]\psi_2(x)+V_{21}(x)\psi_1(x)=0\;\; \text{and} \\\nonumber
\left[H_{33}(x)-E\right]\psi_3(x)+V_{31}(x)\psi_1(x)=0.
\end{eqnarray}
\noindent The above equation after rearranging is given below
\begin{eqnarray}
\left[E-H_{11}(x)\right]\psi_1(x)-V_{12}(x)\psi_2(x)-V_{13}(x)\psi_3(x)=0,\\
\psi_2(x)=\left[E - H_{22}(x)\right]^{-1}V_{21}(x)\psi_1(x)\;\; \text{and} \\
\psi_3(x)= \left[E - H_{33}(x)\right]^{-1}V_{31}(x)\psi_1(x).
\end{eqnarray}
After eliminating both $\psi_2(x)$ and $\psi_3(x)$ from Eq. (97) we get
\begin{equation}
\left(\left[E-H_{11}(x)\right]- V_{12}(x)\left[E - H_{22}(x)\right]^{-1} V_{21}(x) - V_{13}(x)\left[E - H_{33}(x)\right]^{-1}V_{31}(x)\right)\psi_1(x)=0.
\end{equation}
The above equations are true for any general $V_{12}$, $V_{21}$, $V_{13}$ and $V_{31}$. 
The above equation simplify considerably if $V_{12}$, $V_{13}$, $V_{31}$ and $V_{21}$ are Dirac Delta functions,
which we write in operator notation as $V_{12}=V_{21}=K_2 S=K_2 |x_2\rangle \langle x_2|$ and 
$V_{13}=V_{31}=K_3 S=K_3 |x_3\rangle \langle x_3|$.
\noindent The above equation now becomes
\begin{equation}
\left(\left[E-H_{11}(x)\right]-K_2^2\delta (x-x_2)G_2^0(x_2,x_2;E)-K_3^2\delta (x-x_3)G_2^0(x_3,x_3;E)\right)\psi _1(x)=0.
\end{equation}
\noindent This may be written as
\begin{equation}
\left(\left[E-H_{12}(x)\right]-K_3^2\delta (x-x_3)G_3^0(x_3,x_3;E)\right)\psi _1(x)=0,
\end{equation}
\noindent where 
\begin{equation}
H_{12}(x) = H_{11}(x)+K_2^2\delta (x-x_2)G_2^0(x_2,x_2;E).
\end{equation}
For $H_{12}(x)$, one can find the corresponding Green's function $G_{12}(x,x^{\prime };E)$ using the
method as we have used in two state case.
\begin{equation}
\label{156}\left[E - H_{11}\right]\psi _1(x)=K_2^2\delta (x-x_2)G_2^0(x_2,x_2;E)\psi
_1(x),
\end{equation}
\noindent where the right hand side is considered as an inhomogeneous term. The
general solution of this equation can be written as
\begin{equation}
\label{157}\psi _1(x)=\psi _0(x)+\int\limits_{-\infty }^\infty
dxG_1^0(x,x^{\prime };E)K_2^2\delta (x^{\prime }-x_2)G_2^0(x_2,x_2;E)\psi
_1(x^{\prime}),
\end{equation}
\noindent where $\psi _0(x)$ is a solution of the homogeneous equation
\begin{equation}
\label{158}(E - H_{11})\psi _0(x)=0,
\end{equation}
\noindent where
\begin{equation}
\label{159}(E - H_{11})G_1^0(x,x^{\prime };E)=\delta (x-x^{\prime }).
\end{equation}
\noindent So
\begin{equation}
\label{160}\psi_1(x)=\psi _0(x)+K_2^2G_1^0(x,x_2;E)G_2^0(x_2,x_2;E)\psi
_1(x_2).
\end{equation}
\noindent In the above expression
\begin{eqnarray}
\psi _1(x)=\int\limits_{-\infty }^\infty dx^{\prime
}G_{12}(x,x^{\prime };E)\psi _0(x^{\prime })\;\; \text{and}\\\nonumber
\psi _0(x)=\int\limits_{-\infty }^\infty dx^{\prime
}G_1^0(x,x^{\prime };E)\psi _0(x^{\prime }).
\end{eqnarray}
\noindent So Eq. (108) can be written as \newline
$\int\limits_{-\infty }^\infty dx^{\prime }G_{12}(x,x^{\prime };E)\psi
_0(x^{\prime })$=
\begin{equation}
\label{163}\int\limits_{-\infty }^\infty dx^{\prime }G_1^0(x,x^{\prime
};E)\psi _0(x^{\prime
})+K_2^2G_1^0(x,x_2;E)G_2^0(x_2,x_2;E)\int\limits_{-\infty }^\infty
dx^{\prime }G_{12}(x_2,x^{\prime };E)\psi _0(x^{\prime }).
\end{equation}
\noindent The solution in terms of Green's function, extracted from last equation
\begin{equation}
\label{164}G_{12}(x,x^{\prime };E)=G_1^0(x,x^{\prime
};E)+K_2^2G_1^0(x,x_2;E)G_2^0(x_2,x_2;E)G_{12}(x_2,x^{\prime };E).
\end{equation}
\noindent In the above equation we put $x=x_2$ to get
\begin{equation}
\label{165}G_{12}(x_2,x^{\prime };E)=G_1^0(x_2,x^{\prime
};E)+K_2^2G_2^0(x_2,x_2;E)G_1^0(x_2,x_2;E)G_{12}(x_2,x^{\prime };E).
\end{equation}
\noindent So after simplification we get
\begin{equation}
\label{166}G_{12}(x_2,x^{\prime };E)=\frac{G_1^0(x_2,x^{\prime };E)}{%
1-K_2^2G_1^0(x_2,x_2;E)G_2^0(x_2,x_2;E)},
\end{equation}
\noindent so that
\begin{equation}
\label{167}G_{12}(x,x^{\prime };E)=G_1^0(x,x^{\prime };E)+\frac{%
K_2^2G_1^0(x,x_2;E)G_2^0(x_2,x_2;E)G_1^0(x_2,x^{\prime };E)}{%
1-K_2^2G_1^0(x_2,x_2;E)G_2^0(x_2,x_2;E)}
\end{equation}
\noindent Now we will incorporate the effect of third state which is coupled to first state only, i.e. we will solve Eq. (102) in terms of Greens function.
\begin{equation}
\label{170}G_{13}(x,x^{\prime };E)=G_{12}(x,x^{\prime };E)+\frac{%
K_2^2G_{12}(x,x_3;E)G_2^0(x_3,x_3;E)G_{12}(x_3,x^{\prime };E)}{%
1-K_2^2G_{12}(x_3,x_3;E)G_2^0(x_3,x_3;E)}
\end{equation}
\noindent So after incorporating $N$-th state, where all states are coupled to the
first one only we will get
\begin{equation}
\label{171}((E-H_{1[N-1]}(x))-K_N^2\delta (x-x_N)G_N^0(x_N,x_N;E))\psi
_1(x)=0,
\end{equation}
\noindent where
\begin{equation}
\label{172}H_{1[N-1]}(x)=H_{11}(x))+\sum\limits_{n=2}^{N-1}K_n^2\delta
(x-x_n)G_n^0(x_n,x_n;E).
\end{equation}
\noindent In this case, one can find the Green's function $G_{1N}(x,x^{\prime };E)$
using the method as we have already used.
\begin{equation}
\label{173}G_{1N}(x,x^{\prime };E)=G_{1[N-1]}(x,x^{\prime };E)+\frac{%
K_2^2G_{1[N-1]}(x,x_N;E)G_N^0(x_N,x_N;E)G_{1[N-1]}(x_N,x^{\prime };E)}{%
1-K_2^2G_{1[N-1]}(x_N,x_N;E)G_N^0(x_N,x_N;E)}.
\end{equation}
\noindent In this case we can calculate $G_{1N}(x,x_0;E)$ if we know $G_{1(N-1)}(x,x_0;E)$. So using $G_{1N}(x,x_0;E)$, one can calculate wave function explicitely and from the wave function one can easily calculate transition probability from one diabatic potential to all other diabatic potentials.

\section{Conclusions:}

\noindent Starting from few simple examples, we have proposed a general method for finding the exact analytical solution
for the two state quantum scattering problem in presence of a delta function coupling. 
Our solution is quite general and is valid for any potential. We have also extended our model to deal with general one dimensional multi-channel scattering problems. The same procedure is also applicable to the case where $S$ is a non-local operator and may be represented by $S\equiv |f\rangle $ $K_0$ $\langle g|$, where $f$ and $g$ are
arbitrary acceptable functions. Choosing both of them to be Gaussian should
be an improvement over the delta function coupling model. $S$ may even be a linear combination of such operators.

\section{Acknowledgments:}

\noindent It is a pleasure to thank Prof. K. L. Sebastian for suggesting this problem. The author would like to thank Prof. M. S. Child for his kind interest, suggestions and encouragement.

\end{document}